\documentclass{acm_proc_article-sp}

\usepackage[hyphens]{url}
\usepackage[ruled,vlined]{algorithm2e}
\usepackage{url} 
\usepackage{color}

\begin{document}

\title{A Critical Look at Decentralized Personal Data Architectures}
%
%
%
%
%

\numberofauthors{5} 
%
%
%

	\author{
		Arvind Narayanan \\
			{\small \texttt{relax@stanford.edu}}	
		\and
			Solon Barocas\\
			{\small \texttt{solon@nyu.edu}}
		\and
			Vincent Toubiana\\
		{\small \texttt{vincent.toubiana@alcatel-lucent.com}}
		\and
			Helen Nissenbaum \\
			{\small \texttt{hfn1@nyu.edu}}
		\and
			Dan Boneh \\
			{\small \texttt{dabo@cs.stanford.edu}} 
	}
\makeatletter
\def\@copyrightspace{}
\g@addto@macro\@maketitle{
\vspace{-20pt}
\begin{center}
\aufnt\@date
\end{center}
\vspace{20pt}
}
\makeatother
\maketitle
\begin{abstract}

While the Internet was conceived as a decentralized network, the most widely used web applications today tend toward centralization. Control increasingly rests with centralized service providers who, as a consequence, have also amassed unprecedented amounts of data about the behaviors and personalities of individuals.

Developers, regulators, and consumer advocates have looked to alternative decentralized architectures as the natural response to threats posed by these centralized services.  The result has been a great variety of solutions that include personal data stores (PDS), infomediaries, Vendor Relationship Management (VRM) systems, and federated and distributed social networks.  And yet, for all these efforts, decentralized personal data architectures have seen little adoption.

This position paper attempts to account for these failures, challenging the accepted wisdom in the web community on the feasibility and desirability of these approaches. We start with a historical discussion of the development of various categories of decentralized personal data architectures. Then we survey the main ideas to illustrate the common themes among these efforts. We tease apart the design characteristics of these systems from the social values that they (are intended to) promote. We use this understanding to point out numerous drawbacks of the decentralization paradigm, some inherent and others incidental. We end with recommendations for designers of these systems for working towards goals that are achievable, but perhaps more limited in scope and ambition.

\end{abstract}

\section{Brief Historical Overview}

The search for alternatives to centralized aggregation of personal data began in the late 1990s which saw a wave of so-called `negotiated privacy techniques' including commercial `infomediaries' \cite{comingbattle, Dix}. These entities would store consumers' data and help facilitate the drafting of contracts that set the terms of the exchange and use of data.  The 1999 book {\em Net Worth} \cite{networth} galvanized both industry and privacy advocates, generating hopes for a future in which privacy problems could be solved through a mix of decentralized storage and private contracts, potentially obviating the need for privacy law or even the adoption of fair information practices \cite{catlett, dawn}.

Within five years, nearly all of this excitement had faded and all commercial (Persona, Privada, Lumeria, etc.) and community (P3P) initiatives had floundered \cite{ackerman} --- some in truly spectacular fashion, such as AllAdvantage.  And yet, by the end of the decade, many new initiatives and projects that shared almost identical goals emerged.  Vendor Relationship Management (VRM) \cite{Mitchell_Henderson_Searls_2008} has gained steady momentum as a general set of principles that aim simultaneously to improve user privacy, enhance customer autonomy, and increase market efficiency through a combination of mechanisms that aggregate data in a single (per-user) repository under users' control and tools to negotiate agreements that would grant outside organizations access to and use of that data.
 
Parallel efforts to develop so-called personal data stores (PDS), personal data servers, personal data lockers/vaults, and personal clouds \cite{personalcloud} have focused more narrowly on the platforms and protocols to support unified repositories of user data that could be managed locally by the user or outsourced to a trusted third party.  The impetus for these projects are varied, ranging from user interest in aggregating one's own data in a single location to better derive benefits from their mixing and matching to more explicit interests in privacy (user control) and commerce (a market place for sharing, including possibilities for cash payments in exchange for data) \cite{startupcircle}.

The similarities between these and earlier efforts can be quite stark: Mydex's recent white paper, ``The Case for Personal Information Empowerment" \cite{mydex}, recapitulates much that was described in a white paper released a full decade earlier by Lumeria, a failed infomediary \cite{Lumeria}.  To describe this as a simple case of ``an idea whose time has come" would be to miss the important lessons that these earlier and recurring failures should offer those who wish to pursue decentralized personal data architectures.

Decentralized social networking has been a largely parallel, sometimes overlapping line of development with similar motivations. We subdivide such social networks into federated (ecosystem of interoperable implementations in the client-server model) and distributed (peer-to-peer). The term distributed social networking is frequently but incorrectly used to describe all decentralized social networks.

While some early thinking in the semantic web community could be classified in this category,\footnote{The Internet Archive lists a version of the {\em Friend of a Friend} (FOAF) project (www.foaf-project.org) from August 2003, and other efforts may be older.} for the most part decentralized social networking appears not to have anticipated the success of mainstream commercial, centralized social networks, but rather developed as a response to it. Indeed, prominent members of the web community dismissed social networks until 2007--2008 (for example, \cite{atwood} and \cite{winer}) and academic computer scientists appear to have considered it a passing fad as well --- in our survey we see a sharp spike in interest among researchers around this time frame.

A series of well-publicized privacy mishaps by Facebook and Google starting in 2009 that reached its crescendo around the 2010 f8 developer conference stirred up interest among the public and policymakers.\footnote{For an article typifying public opinion during that period, see \cite{rogue}.} Perhaps the most well known project that resulted is Diaspora\footnote{\url{https://joindiaspora.com/}}, which was funded in excess of \$200,000 via the crowd funding platform kickstarter.com. As of this writing Wikipedia lists about 40 decentralized social networks \cite{WP:DSN}, most of which are federated, whereas the academic literature has focused on distributed social networking for natural reasons, since those present more research challenges.

\section{Representative Survey}
\label{sec:RepresentativeSurvey}

Rather than attempt an exhaustive survey, in this section we list the key ideas that have been explored in the course of developing decentralized designs. There has been a great fecundity of creative and complex ideas in this space spanning the realms of technology, law and economics; we are unable to present them in detail due to space constraints. We refer the reader to the cited works.

The core idea of an infomediary is that of a trusted third party that interfaces between the user and commercial entities such as marketers \cite{networth}. Users' personal data can be manually given to the infomediary, as in Lumeria, or collected through passive monitoring, as in AllAdvantage and other systems \cite{givens}.  That information can then be utilized without explicit monetization (Mydex, etc.), or users can be paid for their data (AllAdvantage, Bynamite \cite{bynamite}, etc).  It has variously been argued that telecommunications providers \cite{vodafone, ayres}, banks \cite{SWIFTDAG} and other parties such as providers of home entertainment set-top boxes are ideally suited to play the role of the intermediary.  An infomediary might also enable a targeted \emph{attention market} \cite{AllAdvantage} based on user preferences.

Kang et al. introduce the intriguing idea of \emph{licensing} intermediaries to increase their trustworthiness \cite{Kang_2010}. In the other direction, Vendor Relationship Management systems largely eliminate the infomediary as a separate entity, and instead replace it with a software agent \cite{Mitchell_Henderson_Searls_2008}. Some software intermediaries like Adnostic use cryptography to achieve additional privacy properties \cite{adnostic:privacy}. Other ideas for improving privacy include fine-grained access control lists \cite{mun2010personal}.

Both VRM and infomediary systems often emphasize benefits to the firm from the intermediated nature of the exchange. Goldman \cite{goldman_2006} envisions that software agents  will make marketing messages perfectly relevant, eliminating externalities from wasted attention. By Coase's theorem \cite{coase}, this will lead to a socially optimal level of marketing. 

Turning to social networks, the key challenge of distributed social networks is hosting and message transfer. One solution is to encrypt messages and store them in a distributed hash table \cite{buchegger:peerson, Aiello_Ruffo_2010}. Another is ``social replication'': messages are stored in plaintext in a redundant manner by those who have access rights (typically friends of the message poster) \cite{socialreplication}. Message passing sometimes exploits the relationship between the social graph and the topology of the physical network \cite{socialbutterfly, buchegger:peerson}.

Another frequent goal is keeping edges of the graph secret, for which various solutions have been proposed: a cryptographic approach \cite{backes}, anonymous routing \cite{Cutillo_Molva_Strufe_2009} and friend-to-friend networks such as Freenet in `darknet' mode \cite{Clarke_Miller_Hong_Sandberg_Wiley_2002}. Persona 
takes the cryptographic heavy-lifting a step further to enable fine-grained access control using attribute-based encryption \cite{persona}.

Other models for hosting have been explored. In vis-a-vis, each user owns an EC2 virtual host that is active at all times \cite{Shakimov_Lim_Li_Liu_Varshavsky_2011}, whereas FreedomBox\footnote{\url{http://freedomboxfoundation.org/}} proposes cheap plug computers. Lam et al. have proposed email as a backend \cite{Fischer_Lam_2011} and ephemeral networks on smartphones \cite{Dodson_Lam_2010}. Unhosted\footnote{\url{http://unhosted.org/}} proposes separating data from code, but keeping both in the cloud. Along similar lines, Frenzy\footnote{\url{http://frenzyapp.com/}} is a distributed social network software with Dropbox as the backend. Polaris proposes reducing existing social networks such as Youtube and Twitter to datastores and layering a social network on top, with smartphones providing access control management interfaces \cite{polaris}.

Finally, federated social networks aim to create an ecosystem of standards-based interoperable implementations of social networks. Some designs such as Diaspora are a hybrid between distributed and federated. OStatus, being coordinated by the W3C, represents an interesting approach to standardization for federated microblogging: it references a suite of existing protocols rather than developing them from scratch.

\section{Classification}
\label{sec:Classification}
\begin{table}[ht]
\caption{The four types of architectures that are the subject of our study} 
\label{label:results}
\centering 
\begin{tabular}{ |c | c| c |  }
\hline 
 & Commerce, Health etc. & Social Networking\\
 \hline 
Self-hosted & PDS / VRM	& Distributed\\
\hline
Outsourced & Infomediary & Federated\\
\hline
\end{tabular}
\label{table:2x2} 
\end{table}

We emphasize that the division in Table \ref{table:2x2} is only meant to provide the reader with a rough mental map and is far from precise. The vertical axis, in particular, is closer to a spectrum than a strict division. The terms Personal Data Store and Vendor Relationship Management do not appear to have a single definition. Also, some PDS projects are application-agnostic, but these tend to be software libraries/platforms rather than complete user-facing systems.

Towards a finer-grained classification and understanding of different projects, we propose the following (non-independent) axes that are components of what it means for an architecture to be decentralized. 

\vspace{-5mm}
\begin{enumerate}
	\item  Locus of data hosting: this could be remote (centralized), by a trusted third party (infomediary), distributed (peer-to-peer), or local (i.e., on the user's device). 
  \item Open standards vs. proprietary.
  \item  Open vs. closed-source implementations.
  \item  Data portability: Data export (for users), APIs (for third parties), or none.
\end{enumerate}
\vspace{-2mm}

The above are technical characteristics; one might also try to classify systems in terms of the social values they espouse. We discuss four in particular.

\vspace{-5mm}
\begin{enumerate}
    \item Privacy: According to Nissenbaum \cite{contextualintegritypaper, contextualintegritybook}, systems that attempt to preserve privacy should attempt to preserve the integrity of the context in which actors engage with each other. They should do this by ensuring that information flows respect the norms of the context. To the degree that systems better model and mediate appropriate information flows, they will advance the privacy interests of their users. This view will inform the discussion in Section \ref{sec:oncontrol}.
    \item Utility: We refer to the overall social benefit of the system, in the sense of welfare maximiation in economics. One way to achieve increased utility is through greater interoperability or data portability.
    \item Cost: Cost encompasses hosting and bandwidth costs as well as software development and maintenance costs. Centralized and decentralized systems behave very differently: in the former case there is typically a single entity that bears all the costs whereas in the decentralized setting it can be split among users and various software creators and service providers. Comparing these alternatives may therefore be tricky.
    \item Innovation: We must also consider how quickly different systems are able to evolve and adapt. Some have argued that open standards catalyze innovation while others point to the time and monetary costs of standardization. The strength of the business model, the extent of market competition, the ability to harness and analyze data, and legal compliance requirements are some of the other factors that affect how conducive a system is to innovation.
\end{enumerate}
\vspace{-2mm}

Values may not be immediately deducible from the technical design of a system, but may instead only be observable empirically. Indeed, we suggest that much of the reason for what we see as overenthusiastic claims about decentralized systems is that design characteristics have been confused with values. We discuss two prominent cases in detail in Sections \ref{sec:oncontrol} and \ref{sec:openstandards}. Moreover, we doubt whether any architecture could optimize for all values simultaneously.

\section{Drawbacks of Decentralization}
\label{sec:Drawbacks}

In this section we present some underappreciated drawbacks of decentralized architectures. Not all of these apply to all types of systems, nor is any of them individually a decisive factor. But collectively they may help explain why decentralization faces a steep road ahead, and why even if adopted, decentralization will not necessarily provide all the benefits that its proponents believe will automatically flow from it.

An architecture without a single point of data aggregation, management and control has several {\bf technical} disadvantages. First is functionality: there are several types of computations that are hard or impossible without a unified view of the data. Detection of fraud and spam, search, collaborative filtering, identification of trending topics and other types of analytics are all examples. Decentralized systems also suffer from inherently higher network unreliability, resulting in a tradeoff between consistency and availability (formalized as the CAP theorem \cite{WP:CAP}); they may also be slower from the user's point of view.\footnote{Google reports that users exposed to an additional delay of as little as 100ms performed a statistically significantly smaller number of searches \cite{speedmatters}.} The need for synchronized clocks and minimizing data duplication are other challenges.

The benefits and costs of standardization are a prominent socio-technical factor. Many decentralized systems depend on multiple interoperating pieces of software, which requires standardization of technical protocols, design decisions, etc. On the one hand, such an ecosystem could promote long-term innovation; on the other hand, these processes (e.g., HTML5) move at a far slower pace than Facebook or an ad network which can roll out features over the timespan of days or weeks. Shapiro notes two benefits of standardization: greater realization of network effects and protection of buyers from stranding, and one cost: constraints on variety and innovation, and argues that the impact on competition can be either a benefit or a cost \cite{shapiro}. 

Let us now turn to {\bf economics}. Centralized systems have significant \emph{economies of scale} which encompasses hosting costs, development costs and maintenance costs (e.g., combating malware and spam),\footnote{Facebook has built a highly sophisticated real time ``immune system'' which relies in part on human operators \cite{FIS}.} branding and advertising \cite{peles}. A related point in the context of social networks: we hypothesize that the network effect is stronger for centralized systems due to tighter integration.

\emph{Path dependence} is another key economic issue: even if we assume that centralized and decentralized architectures represent equally viable equilibria, which one is actually reached might be entirely a consequence of historical accident. Most of the systems under our purview -- unlike, say, email -- were initially envisioned as commercial applications operating under central control, and it is unsurprising they have stayed that way.

The theory of \emph{unraveling} suggests that infomediaries in particular might {\em not} in fact represent a stable equilibrium. For an infomediary to succeed, consumers and businesses must transact through the intermediary rather than directly with each other. But either side of this market might see participants iteratively defecting, resulting in unraveling of the market. Chen et al. discuss how this might happen from the businesses' side \cite{chen}, and Peppet discusses it from the consumer side \cite{Peppet_2010}. However, it is not fully clear why many types of intermediaries have taken hold in many other markets --- employment agents, goods appraisers, etc --- but not in the market for personal data.

A variety of \textbf{cognitive} factors hinder adoption of decentralized systems as well. First, the fact that decentralized systems typically require software installation is a significant barrier. Second, more control over personal data almost inevitably translates to more decisions, which leads to cognitive overload. Third, since users lack expertise in software configuration, security vulnerabilities may result. A related point is that users may be unable to meaningfully verify privacy guarantees provided through cryptography.

Finally, we find that decentralized social networking systems in particular fare poorly in terms of mapping the norms of information flow. Access control provides a very limited conception of privacy. We provide several examples. First is the idea of ``degrees of publicness.'' For example, on Facebook a post may be publicly visible, yet the site has defenses to stop crawlers which prevents the post ending up in a search engine cache, so that the user may meaningfully hide or delete the post later if they so choose. Second, in current social networks privacy is achieved not only through technical defenses but also through ``nudges'' \cite{privacynudges}.
When there are multiple software implementations, users cannot rely on their friends' software providing these nudges. Third, distributed social networks reveal users' content consumption to their peers who host the content\footnote{This is a particularly serious problem for systems like Contrail \cite{contrail}.} (unless they have a ``push'' architecture where users always download accessible content, whether they view it or not, which is highly inefficient.) Finally, decentralized social networks make reputation management and ``privacy through obscurity'' (in the sense of \cite{Hartzog_2010}) harder, due to factors such as the difficulty of preventing public, federated data from showing up in search results.

\subsection{On Control over Personal Data}
\label{sec:oncontrol}

We now discuss two drawbacks in detail to illustrate the difference between architectural decisions and social values that they are often implicitly assumed to promote. The first is the distinction between control over hosting and privacy. To elucidate this we present a thought experiment.

What does it mean for users to truly host and control their personal data, while still being able to participate in activities such as social networking and personalized commerce? Compared to using Facebook, hosting one's data on a personal EC2 instance certainly puts the user in greater control, but Amazon will turn over user data in response to a subpoena or court order \cite{AWSagreement}.

For any hope of absolute control, users must, at a minimum, host data on their own device resident on their physical property. This is already considerably at odds with the reality of today's consumer Internet: bandwidth to the home is often asymmetric, or connectivity is restricted in other ways (NATs, firewalls), and few individuals possess always-on devices capable of running web services.\footnote{It remains to be seen if smartphones will become practical for this use-case.}

Furthermore, the software running the services must be open-source, and be audited by third-party certification authorities, or by ``the crowd". Silent auto-updates, which is the model that client-side software is increasingly moving towards, would be difficult due to the auditing requirement, perhaps prohibitively so.

Further still, {\em hardware} might have backdoors, and therefore needs an independent trust mechanism as well. The user also needs the time and knowhow to configure redundant backups, manage software security, etc. Finally almost all decentralized architectures face the the problem of ``downstream abuse" which is that the user has no technical means to exercise control over use and retransmission of data once it has been shared \cite{privacyproperty}. 

This thought experiment shows that absolute control is impossible in practice. Further, it suggests that control over information is probably not the right conceptualization of privacy, if privacy is the end goal.

\subsection{Open standards and Interoperability}
\label{sec:openstandards}
Interoperability is a laudable goal; it could enhance social utility, as we have mentioned earlier. However, it has frequently been reduced to the notion of open standards. We argue here that while open standards are a prerequisite for interoperability, there is a big gap between the two. In particular, the efforts at federated social networking all follow open standards, but their actual interoperability status in practice appears to be poor \cite{SWAT0}. Let us examine why this is the case.

One major impediment is that there are too many standards to choose from. For the most basic, foundational component --- identity --- there are many choices: OpenID, WebID and others. While it is possible to connect these to each other, it requires extra effort.  As we get to more complex (but still basic) functionality such as federation of messages, we find on the one hand Atom/PubSubHubbub etc. and the OStatus suite\footnote{\url{http://ostatus.org/}} on top of it, and on the other hand XMPP and the Wave federation protocol\footnote{\url{http://www.waveprotocol.org/}} on top of it. It appears that the former is gradually winning out, but this is a slow process.

The second major impediment is that as soon as we get past the basics like identity, friendship and status updates, there is an incredible array of parameters to nail down. Take the apparently trivial issue of what formatting is allowed in a status update. Unless two systems agree on the same standard, they are not interoperable because users of one will see malformatted messages originating from the other. Needless to say, centralized platforms have a large and ever-increasing set of features --- photos, video chat, polls, to name a few --- all of which would require standardization in the federated context.  Finally, access control in a federated setting presents hard technical challenges.

The practical upshot is that the only suite of standards that shows any signs of meaningful interoperability is StatusNet\footnote{\url{http://status.net/}} --- microblogging is both text based, largely eliminating the formatting issue, and typically public, sidestepping the access-control issue ---  although identi.ca remains the only implementation with meaningful adoption. Even though this system limits status updates to text, a version of the formatting problem still plagues it: identi.ca restricts updates to 140 characters in an attempt to maintain some interoperability with Twitter!

We conclude that while federated social networks have the potential to converge on a reasonably interoperable collection of software --- subject to the caveats of differing feature sets and parameters --- it is not simply a matter of making some technical decisions, but instead needs serious developer commitment as well as the involvement of standards bodies with significant authority.

\section{Recommendations}
\label{sec:Recommendations}

Based on our analysis above, we offer the following recommendations for developers of decentralized systems.
\vspace{-5mm}
\begin{enumerate}
	\item Consider the economic feasibility of your design. In particular, are there entities with the economic incentive to play the various roles that are called for? This has perhaps been the most common reason for the lack of adoption of past proposals and projects.
  \item Pay heed to conceptual fidelity. Are you shooting at the right target? Do people have the values you think they do? Do they really want the features/benefits you claim they want? As one example, there have been a multiple of projects that attempt encrypted communication over Facebook and other social networks (NOYB \cite{noyb}, FlyByNight \cite{flybynight}, Lockr \cite{lockr}, FaceCloak \cite{facecloak}, Scramble! \cite{scramble}, etc.), but the lack of adoption suggests that the usability costs do not outweigh the benefits to users.
  \item Incorporate other notions of regulability \cite{Zittrain, lessigcode}. Many decentralized systems represent an extreme choice: they seek to achieve privacy and other properties purely through technology, ignoring socio-legal approaches. This extreme may not be optimal. 
  \item Offer advantages other than privacy to users. Privacy is always a secondary feature --- while it might tip the balance between two competing products, users rarely pick a product based on privacy alone. For example, distributed social networking can enable some location-specific functionalities through peer-to-peer networking even when there is no Internet access.
  \item Design with standardization in mind. One of the disadvantages we have identified is the proliferation of non-interoperable systems. Open standards are not enough: developers must actively prioritize interoperability and write and maintain glue code to interface with other systems.
  \item Target limited feature sets. A system like Facebook is a large, complex moving target. Attempting to create a decentralized version of it is a futile endeavor. Instead, systems that embody the `minimum viable product' strategy might succeed better in the market. Decentralized microblogging appears to be a relatively attainable goal at the present time, and censorship resistance is a goal for which there is much demand.
  \item Work with regulators. As numerous law/economics scholars have pointed out, market solutions appear to underprovide privacy and regulation can help tweak the environment to address this imbalance \cite{Schwartz}. Those who wish to see the personal data ecosystem flourish would do well to support regulatory interventions such as transparency and opt-out that can help level the playing field between centralized and decentralized systems.
\end{enumerate}

\section{Conclusion}
\label{sec:conclusion}

In this position paper we have taken a look back at the efforts to build decentralized personal data architectures motivated either by discontent with the status quo, or as a better way to organize information markets and leverage new commercial opportunities, or a combination of both. We hope we have provided some mental clarity to the reader on the similarities, differences and common themes between the various systems and brought fresh perspective to the question of why they have largely floundered.

We hope to kick off a more tempered discussion of the future of personal data architectures in both scholarly and hobbyist/entrepreneurial circles, one that is informed by the lessons of history. There is much work to be done along these lines --- application of economic theory can shed light on questions such as the relative strength of network effects in centralized vs. decentralized systems. Empirical methodology such as user and developer interviews would also be tremendously valuable. While we have provided some suggestions for developers, in the future we hope to identify specific application domains that are relatively amenable to the adoption of decentralized architectures, as well as to provide concrete recommendations for policymakers who might wish to foster a different market equilibrium.

{\bf Acknowledgement.} The first author would like to thank Monica Lam and the other members of the MobiSocial project for enlightening discussions, Deirdre Mulligan, Nick Doty and Jennifer King for helping develop ideas on a multi-factor approach to privacy, Alejandro Molnar for general education about economics, Alessandro Acquisti for sharing his bibliography on the economics of privacy and numerous online commenters for perspectives, links and information.

\newpage
\onecolumn
\bibliographystyle{abbrv}
\bibliography{dumw}
\end{document}